\documentstyle[12pt]{article}
\def\tw{\theta_{\!\rm w\,}}
\def\sintw{\sin\tw}
\def\costw{\cos\tw}
\def\bm#1{{\mbox{\boldmath $#1$}}}
\newcommand{\phat}[1]{\hat{\phi}^{#1}}

\begin{document}
\pagestyle{empty}
\pagenumbering{arabic}
\begin{flushright}
\baselineskip=16pt
{\footnotesize DAMTP--98-2}\\
{\footnotesize FERMILAB--Conf-98/018-A}\\
{\footnotesize hep-ph/9801286}
\end{flushright}
\renewcommand{\thefootnote}{\fnsymbol{footnote}}
\footnotesep=14pt
\vspace{0.15in}
\baselineskip=24pt
\begin{center}
{\Large\bf Magnetic Fields from the}\\
{\Large\bf Electroweak Phase
Transition\footnote{\baselineskip=14pt\noindent
Talk presented at the International Workshop on Particle
Physics and the Early Universe \mbox{(COSMO--97)} in Ambleside,
England, 15-19 Sept 1997. To appear in the
proceedings (World Scientific).}}\\
\baselineskip=16pt
\vspace{0.75cm}
\addtocounter{footnote}{1}
{\bf Ola T\"{o}rnkvist\footnote{Electronic address:
{\tt O.Tornkvist@damtp.cam.ac.uk}}}\\ {
{\small\em NASA/Fermilab Astrophysics Center}\\
\baselineskip=14pt
{\small\em MS-209, P.O.~Box 500, Batavia, IL
60510-0500, U.S.A.}\\
{\small\em and}\\
{\small\em Department of Applied Mathematics and Theoretical Physics}\\
{\small\em University of
Cambridge, Cambridge CB3 9EW,
United Kingdom\/\footnote{Present address.}}\\}
\baselineskip=16pt
\vspace*{.75cm}
{1 January 1998}
\end{center}
\vspace{0.2cm}
\begin{abstract}
I review some of the mechanisms through which primordial
magnetic fields may be created in the electroweak phase
transition. I show that no magnetic fields are produced
initially from two-bubble collisions in a first-order
transition. The initial field produced in a
three-bubble collision is computed. The evolution of
fields at later times is discussed.
\end{abstract}
\newpage
\pagestyle{plain}
\renewcommand{\thefootnote}{\arabic{footnote}}
\setcounter{footnote}{0}
\baselineskip=16pt

\section{Introduction}
Observations, of e.g.\ the Faraday rotation of the polarisation of
light from distant sources as it passes through galaxies, reveal
that galaxies typically possess a magnetic field with strength
of the order of $10^{-6}$ Gauss. The direction of
the field appears to be
correlated with each galaxy's axis of rotation, which suggests
that it was generated during the epoch of galaxy formation
through magnetohydrodynamic (MHD) processes. Such
processes
include dynamo mechanisms, which may amplify the magnetic field by many
orders of magnitude, and
nonlinear inverse-cascade mechanisms, which increase the correlation
length by transferring power to long-wavelength modes \cite{xqs}.

Because the magnetic field $\bm{B}$ enters homogeneously in the
MHD equations, an initial seed field is needed,
which must be
larger than about $10^{-20}$ Gauss.
The most natural
origin of such a seed field is the electroweak phase transition
(EWPT), as that
is the earliest time
a magnetic field can
come into
existence as a gauge-invariant physical quantity. The initial
field is very strong, $B_0\sim M_W^2\sim 10^{24}$ Gauss, and
has a correlation length $\xi\sim M_W^{-1}$ . As the universe
expands, the average magnetic field decreases as the inverse
square of the scale factor, $B(t) \propto [a(t)]^{-2}$,
since in the conducting plasma the magnetic flux through surfaces
bounded by comoving closed curves is conserved. Silk damping,
the diffusion of photons in the charged plasma, may further reduce
the field.

A seed field can arise in the EWPT in several interesting ways \cite{xqs}.
If the transition is first-order, the field may result
{}from the turbulent motion of charged plasma layers on expanding
bubble walls, or in the collision of bubbles that contain
different phases
of the Higgs field \cite{KibVil,SafCop}.
Also in a second-order transition the seed field
may emerge spontaneously in the EWPT, as proposed by
Vachaspati \cite{Vacha},
through the process of symmetry breaking itself.

In this talk, I will concentrate on magnetic-field generation in
collisions of electroweak bubbles. For this purpose, it is necessary
first to reexamine the definition of the electromagnetic field in
the broken-symmetry phase.

\section{Gauge-invariant definition of the electromagnetic field}
 In the unitary gauge, $\Phi=(0,\rho)^{\top}$ with
$\rho\equiv {\rm constant}$,
the electromagnetic field tensor takes the familiar form
${\cal F}_{\mu\nu}^{\rm em} =
\partial_\mu A_\nu - \partial_\nu A_\mu$,
where $A_\mu = \sintw W^3_\mu + \costw Y_\mu$ and $W^3_\mu,Y_\mu$
are defined by their occurrence in the covariant derivative of the
Higgs field, ${\cal D}_\mu\Phi=
(\partial_\mu - i g  W^a_\mu \tau^a/2 - ig'Y_\mu/2)\Phi$.
Here $A_\mu$, by construction, does not couple to the Higgs field $\Phi$;
it represents the massless photon.

 For a general Higgs field
$\Phi=(\phi_1(x),\phi_2(x))^{\top}$, the field $A_\mu$ as defined above
couples to $\phi_1(x)$ and becomes massive. Evidently, at the
position $x$, $A_\mu$ is no longer the photon field, and
the curl of $A_\mu$ cannot be interpreted as the electromagnetic
field tensor. Instead, one must find a gauge-invariant definition
of ${\cal F}_{\mu\nu}^{\rm em}$
whose value in any gauge coincides with that in the unitary gauge.
Such a definition was proposed in Ref.~\cite{me}. In terms of the
three-component unit isovector $\hat{\phi}^a=
(\Phi^{\dagger}\tau^a\Phi)/(\Phi^{\dagger}\Phi)$ it is given by
\begin{equation}
\label{Fem}
{\cal F}^{\rm em}_{\mu\nu} :=
-\sintw \phat{a} F^a_{\mu\nu} + \costw F^Y_{\mu\nu}
+ \frac{\sintw}{g} \epsilon^{abc} \phat{a} (D_\mu\phat{})^b
(D_\nu\phat{})^c~,
\end{equation}
where $(D_\mu\phat{})^a=\partial_\mu\phat{a} + g\epsilon^{abc}
W^b_\mu\phat{c}$.
The last term in Eq.~(\ref{Fem}) correctly takes into account
electromagnetic fields associated with the phases of the Higgs field.
Unlike previous
definitions, Eq.~(\ref{Fem}) eliminates contributions from
neutral currents and subtracts out the $W$-boson magnetic
dipole-moment density also when the magnitude
$(\Phi^{\dagger}\Phi)^{1/2}$ has a spatial dependence \cite{me}.
In addition, it satisfies
the Bianchi identity $\epsilon^{\mu\nu\alpha\beta} \partial_\nu
{\cal F}^{\rm em}_{\alpha\beta}=0$, which ensures that there is
no magnetic charge or magnetic current in the absence of magnetic
monopoles.

\section{Magnetic fields from bubble collisions}
In a first-order EWPT magnetic fields can be created in the
collision of expanding bubbles that contain
different phases of the Higgs field. This was first investigated
in the abelian U(1) model \cite{KibVil}. Because of the phase gradient,
a gauge-invariant current $j_k=i q [\phi^{\dagger} D_k \phi -
(D_k\phi)^{\dagger} \phi]$ develops across the surface of
intersection of
the two bubbles, where $D_k=\partial_k-i q V_k$. The current
gives rise to a ring-like flux of the field strength $\partial_i V_j
-\partial_j V_i$ which takes the shape of a girdle encircling
the bubble intersection region.

This mechanism can be generalised to the electroweak SU(2)$\times$U(1)
theory, where the initial Higgs field in the two
bubbles is of the form
\begin{equation}
\label{higgs}
\Phi_1=\left(\begin{array}{c} 0\\ \rho_1(x)
\end{array}\right)~,
\quad\quad \Phi_2 = \exp(i\frac{\theta}{2} n^a\tau^a)
\left(\begin{array}{c} 0\\ \rho_2(x)\end{array}
\right)
\end{equation}
and
the Higgs
phases have equilibrated to constant values inside each
bubble.
Saffin and Copeland \cite{SafCop} found that such an initial
 configuration can be
written globally in the same form as $\Phi_2$, provided that
$\theta\to\theta(x)$ and $\rho_2(x)$ is
replaced by a different function $\rho_3(x)$. One may also assume that
the initial configuration has zero field strength, $F^a_{\mu\nu}=
F^Y_{\mu\nu}=0$. Then, except in the pathological case of singular
vacuum configurations, one may choose also the gauge potentials
to be zero, $W^a_\mu=Y_\mu=0$.

For $\Phi=\Phi_1$ one obtains $\bm{\hat{\phi}}=\bm{\hat{\phi}}_0 \equiv
(0,0,-1)$. It is
easy to show that, as $\Phi$ interpolates from $\Phi_1$ to $\Phi_2$,
the isovector $\bm{\hat{\phi}}$ is rotated by an angle $\theta$
about the axis $\bm{n}$. Saffin and Copeland discovered that, in the
cases where $\bm{n}$ is parallel or perpendicular to
$\bm{\hat{\phi}}_0$,
the dynamics reduces to the U(1) problem already solved \cite{SafCop}.
The ring-like flux in these two
cases corresponds to the gauge field of an electroweak
$Z$-string or $W$-string, respectively.

The issue of magnetic-field creation in bubble collisions
has so far not been properly addressed in the electroweak theory.
Because the field tensors and gauge potentials are initially zero,
the only contribution to the magnetic field
comes from the last term in Eq.~(\ref{Fem}), which reduces to
$\sintw  \bm{\hat{\phi}}\cdot(\partial_\mu\bm{\hat{\phi}}\times
\partial_\nu\bm{\hat{\phi}})/g$. Because $\bm{\hat{\phi}}$
is obtained by rotating $\bm{\hat{\phi}}_0$  by an angle
$\theta$ about $\bm{n}$,
one finds that
$\partial_\mu\bm{\hat{\phi}} = \partial_\mu\theta\,
(\bm{n}\times\bm{\hat{\phi}})$. It then follows that neither
electromagnetic
fields nor electric currents are present initially in the
collision \cite{me}.
I emphasise that this result is
very different from the U(1) case, where a field strength was
present from the instant of collision of the bubbles.

Let us now consider the subsequent field evolution. Because
${\cal F}^{\rm em}_{\mu\nu}$ satisfies the Bianchi identity,
$\partial\bm{B}/\partial t= - \nabla\times\bm{E} = 0$ initially.
Likewise, since the electric current vanishes initially, if follows
that $\partial\bm{E}/\partial t= \nabla\times\bm{B} = 0$
initially. Non-zero $\bm{E}$ and $\bm{B}$ can thus arise only
in ``second order'' through later evolution of the Higgs
field that renders the last term in Eq.~(\ref{Fem}) nonzero.

As we have seen, no magnetic field is produced initially in the
collision of two bubbles in the EWPT. One can
show that the reason is
that the unit vector $\bm{n}$ is a constant. In contrast, in a
three-bubble collision $\bm{n}$ will have a spatial dependence,
and
a magnetic field will appear already at the instant of collision.
If three bubbles are needed, however, the importance of bubble
collisions for magnetic-field generation is diminished, as the
third bubble must impinge before the phases of the first two have
equilibrated.
The initial configuration of three bubbles can
be written globally as
$\Phi=\exp [i\,(f(x) m_a + g(x)n_a)\tau^a]~(0,\rho(x))^{\top}$,
where the constant unit vectors $\bm{m}$ and $\bm{n}$
are not collinear.
Defining $R\equiv \sqrt{f^2+g^2}$ and taking,
for simplicity, $\bm{m}\perp\bm{n}$, we obtain
\begin{equation}
\label{threebub}
{\cal F}^{\rm em}_{\mu\nu} =
\frac{4\sintw}{g}  f_{[,\mu}g_{,\nu]}
\left(-\frac{\sin  2R}{2R} (m_1 n_2-m_2 n_1) +
\frac{\sin^2 R}{R^2} (f n_3 - g m_3) \right)~,
\end{equation}
which yields a nonzero magnetic field
as long as $\nabla f\times\nabla g\neq 0$.

\subsection*{Acknowledgments}
I am grateful for support from the COSMO-97 conference, from
EPSRC under Grant GR/K50641 and from
 DOE and NASA under Grant NAG5-2788.

\bibliographystyle{unsrt}

\begin{thebibliography}{1}
\setlength{\parsep}{0pt}
\setlength{\parskip}{0pt}
\bibitem{xqs} For lack of space here, please see references given in
[5] below.
\bibitem{KibVil} T.W.B.\ Kibble and A.\ Vilenkin,
{\em Phys.\ Rev.\ }{\bf D52}, 679 (1995);
P.M.\ Saffin and E.J.\ Copeland,
{\em Phys.\ Rev.\ }{\bf D54}, 6088 (1996);
J.\ Ahonen and K.\ Enqvist,
{\em Magnetic field generation in first order phase transition
bubble collisions\/}, hep-ph/9704334, to appear in
{\em Phys.\ Rev.\ }{\bf D}.
\bibitem{SafCop}P.M.\ Saffin and E.J.\ Copeland,
{\em Phys.\ Rev.\ }{\bf D56}, 1215 (1997).
\bibitem{Vacha}T.Vachaspati,
{\em Phys.\ Lett.\ }{\bf B265}, 258 (1991).
\bibitem{me} Ola T\"{o}rnkvist,
{\em On the Electroweak Origin of Cosmological
Magnetic Fields\/}, hep-ph/9707513, FERMILAB preprint Pub-97/257-A,
 submitted to {\em Phys.\ Rev.\ }{\bf D}.

\end{thebibliography}

\end{document}